\documentclass[a4paper,10pt]{article}
\usepackage{amsmath,amssymb,mathtools}  
\usepackage{soul}                       
\usepackage[usenames,dvipsnames]{xcolor}
\usepackage[raggedright]{titlesec}
\usepackage{scrextend}
\deffootnote{2em}{0em}{\thefootnotemark\quad}
\usepackage{multicol}
\usepackage[left=0.75in,right=0.75in,top=1.0in,bottom=1.25in]{geometry}
\usepackage{hyperref}
\usepackage{xcolor}

\hypersetup{
  colorlinks   = true, 
  urlcolor     = blue, 
  linkcolor    = blue, 
  citecolor   = blue 
}

\newcommand{\dd}{\text{d}}

\setstcolor{red}    
\setulcolor{red}    
\allowdisplaybreaks 

\let\oldabstract\abstract
\let\oldendabstract\endabstract
\makeatletter
\renewenvironment{abstract}
{%
               {\list{}{\addtolength{\leftmargin}{4em} 
                        \listparindent 1.5em%
                        \itemindent    \listparindent%
                        \rightmargin   \leftmargin%
                        \parsep        \z@ \@plus\p@}%
                \item\relax}%
               {\endlist}%
\oldabstract}
{\oldendabstract}
\makeatother

\title{\LARGE\textbf{\textsf{On the Disformal Transformation\\of the Einstein-Hilbert Action}}}

\author{\normalsize Allan L. Alinea${}^{(a),(b)} $\footnote{AL Alinea is no longer affiliated with Osaka University but part of this work was done during his last years in the said university.}}

\date{}

\begin{document}

\maketitle
\vspace{-2.25em}
\noindent
\begin{center}
{\small $ {}^{(a)} $Institute of Mathematical Sciences and Physics,\\ University of the Philippines Los Ba\~nos,\\College, Los Ba\~nos, Laguna 4031 Philippines, alalinea@up.edu.ph}
\\[0.5em]
{\small $ {}^{(b)} $Department of Physics, Osaka University\\
Toyonaka, Osaka 560-0043, Japan}
\end{center}

\bigskip
\begin{abstract}
	\noindent{Disformal transformation is a generalisation of the well-known conformal transformation commonly elaborated in mainstream graduate texts in gravity (relativity) and modern cosmology. This transformation is one of the most important mathematical operations in scalar tensor theories attempting to address pressing problems involving dark energy and dark matter. With this topic yet to penetrate these texts, we present a pedagogically oriented derivation of the disformal transformation of the Einstein-Hilbert action. Along the way of calculation, we encounter apparently problematic terms that could be construed as leading to equations of motion that go beyond second order in derivatives, signalling instability. We demonstrate that these terms can be eliminated and absorbed through the definition of the Riemann curvature tensor. The transformed Einstein-Hilbert action turns out to be a special case of the Horndeski action and the equations of motion for the scalar field that it describes are all up to second order only in derivatives, implying stability.}
\bigskip
\\
\noindent
{\small\textbf{keywords:} \textit{disformal transformation, Einstein-Hilbert action, Horndeski action,\\\phantom{keywordssp} Ostrogradsky theorem}}	
\end{abstract}

\bigskip
\begin{multicols}{2}
\section{Introduction}
\label{secIntro}
The conformal transformation of the metric $ g_{\mu \nu } $ involves the change $ g_{\mu \nu } \rightarrow \widehat g_{\mu \nu } = A g_{\mu \nu }$, where the hat indicates the new metric, $ A = A(\phi )$ is the conformal factor, and $ \phi  $ is some scalar field. It is a scaling transformation that locally preserves angles between curves in the spacetime that the metric describes; the infinitesimal light cones are invariant. The topic of conformal transformation is commonly found in graduate texts in relativity and modern cosmology; see for instance, Refs. \cite{Wald:1984rg,Carroll:2004st,Straumann:2013spu,Mukhanov:2005sc,Weinberg:2008zzc,Peter:2013avv}. The discussion in these texts usually involves the transformations, in sequence, from the Christoffel symbol, Riemann tensor, Ricci tensor, and scalar curvature, to the Einstein-Hilbert action. Owing to the fact that the transformed Einstein-Hilbert action is a new action involving a nonminimally coupled scalar field to gravity, conformal transformation finds its significance in relating nonminimally theories to the Einstein-Hilbert action. In quantum cosmology, such relationship is extended to the gauge-invariant primordial cosmological perturbations---the quantum seeds that gave way to galaxies and clusters of galaxies that constitute the visible Universe. These perturbations are invariant under conformal transformation \cite{Fakir:1990eg,Makino:1991sg,Kubota:2011re}, including their regularized power spectrum \cite{Alinea:2016qlf,Alinea:2017ncx} within the context of inflationary cosmology \cite{Guth:1980zm,Dodelson:2003ft}.

The concept of disformal transformation \cite{Bekenstein:1992pj} was introduced by Bekenstein to relate geometries of the same gravitational theory. It is a generalization of the conformal transformation. Given a metric $ g_{\mu \nu } $, disformal transformation takes the form\footnote{This is certainly not the most general form of disformal transformation. But in this work, we focus our attention to this special case and simply call it disformal transformation.} given by
\begin{align}
	\label{disftransf}
	g_{\mu \nu }
	\quad\rightarrow\quad
	\widehat g_{\mu \nu }
	=
	A(\phi )g_{\mu \nu }
	+
	B(\phi )\phi _{;\mu }\phi _{;\nu },
\end{align}
where the semicolon denotes a shorthand for covariant differentiation\footnote{Obviously, $ \phi _{;\mu } = \partial _\mu \phi $. We find it convenient to write $ \phi _{;\mu } $ in preparation for terms such as $ \phi _{;\mu \nu } $.}; \textit{i.e.}, $ \nabla_{\mu }\phi = \phi _{;\mu } $. The functional $ B(\phi ) $ is the disformal factor. Needless to say, when $ B = 0 $, the transformation reduces to the usual conformal transformation. Being a more general transformation, it introduces a richer set of possibilities for the transformed metric. For instance, disformal transformation, in general, does not locally preserve angles, unlike that of the conformal transformation. The light cones may narrow depending on the form of $ B(\phi ) $. At times, some of these possibilities may be unphysical; e.g., flipping of sign of the metric. In such cases, $ A $ and $ B $ have to subject to some constraints; see the next section.

Nowadays, we know that the sphere of importance of disformal transformation goes way beyond the first call in Ref. \cite{Bekenstein:1992pj}. To cite a few, it is a symmetry transformation of the massless Klein-Gordon equation \cite{Falciano:2011rf}, the Maxwell's equations \cite{Goulart:2013laa}, and the Dirac equation under the Inomata's condition \cite{Bittencourt:2015ypa}. Furthermore, the gauge-invariant primordial cosmological perturbations are invariant under disformal transformation. The symmetry of these perturbations are extended from conformal transformation to disformal transformation within the context of the Horndeski theory (see Refs. \cite{Minamitsuji:2014waa,Tsujikawa:2014uza,Domenech:2015hka,Motohashi:2015pra,Alinea:2020laa}). In the study of black holes in the context scalar tensor theories, disformal transformation can be used to investigate regions of the solution space and their symmetry \cite{BenAchour:2019fdf}. Within the same framework, disformal transformation plays a significant role in the generalization of the Einstein-Hilbert action to possibly address pressing problems about dark energy \cite{Harvey:2009zza,Sakstein:2014isa} and dark matter \cite{Sanders:1996wk}.  

In spite of these, disformal transformation is yet to penetrate mainstream graduate texts on modern cosmology and gravity. In scientific literature, result on disformal transformation for the scalar curvature, for instance, is often quoted without the footprints of derivation. Certainly, the use of Computer Algebra System with suitable tensor package can make the calculation fast and ``easy''. However, we are of the opinion that appropriate train of logic should be understood first before simply using a calculator to add numbers. It is in this spirit that we put forth our first objective in writing this paper, namely, to present a pedagogically oriented derivation of the Einstein-Hilbert action suitable for graduate students. We hope to present a sufficiently detailed calculation, following effectively the same logic as in the respected graduate texts mentioned above. This time however, instead of conformal transformation, we deal with the more complicated disformal transformation, from the metric to the transformed scalar curvature, which in turn, effectively leads to the transformed Einstein-Hilbert action.

Our second objective is about the nature of the transformed action; in particular, its stability for the field $ \phi $. As we shall see, the disformal transformation of the Einstein-Hilbert action results in some ``curious'' terms that could possibly lead to equations of motion beyond second order in derivatives. Based on the Ostrogradsky theorem, this could imply that the associated energy is not bounded from  below; therefore, the system that the action describes could be unstable. We wish to demonstrate that the apparently problematic terms can be made to cancel or be absorbed through the Riemann curvature tensor. This results in disformally transformed Einstein-Hilbert action with corresponding equations of motion for the field $ \phi  $ that are second order in the derivatives. In other words, the action describes a stable scalar-tensor theory for $ \phi  $.

This paper is organized as follows. In the following section, Sec. \ref{secScaCurv}, we present a pedagogically oriented derivation of the disformal transformation of several quantities including the Christoffel symbol, Ricci tensor, and Ricci scalar. In principle, once the calculation for the transformations of the scalar curvature and the metric determinant is done, we can simply write out the transformed Einstein-Hilbert action. But as mentioned above, there are problematic terms that pop out along the way of computation. We deal with these terms in Sec. \ref{secConnHornd}. At the end of this section, we explain that the resulting Einstein-Hilbert action describing the field $ \phi $ is stable. We state our concluding remarks in the last section.

\section{Transformation of the Scalar Curvature}
\label{secScaCurv}
\begin{figure}[htb!]
    \centering
    \includegraphics[scale=1.0]{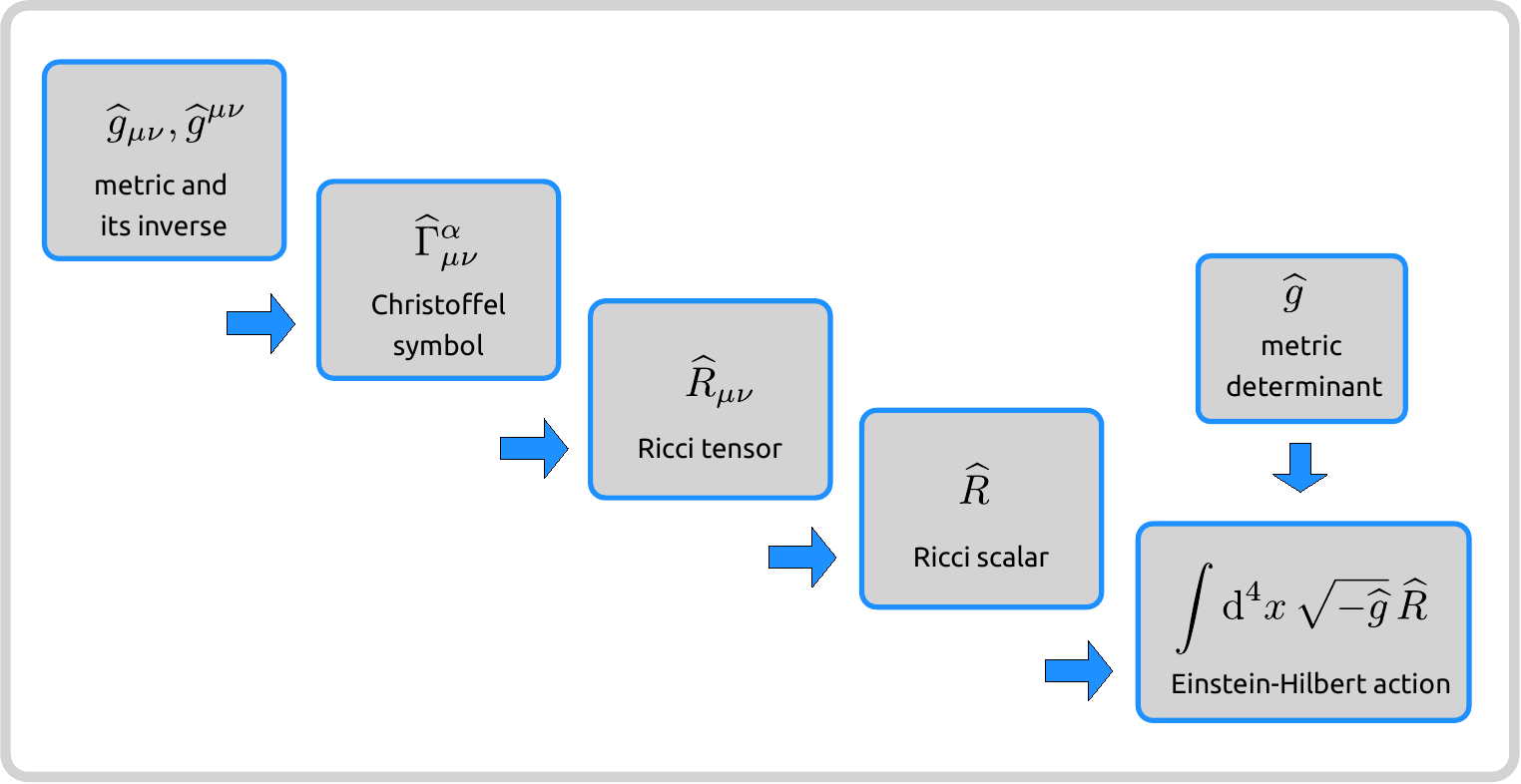}
    \caption{\it Flow chart of the derivation for the disformal transformation of the Eintein-Hilbert action.}
    \label{figEHFlow}
\end{figure}

The Einstein-Hilbert action consists of the integral measure involving the metric determinant and the Ricci scalar. Knowing the metric disformal transformation, one can easily find the determinant. Much of the calculations then involve finding the disformally transformed Ricci scalar. Figure \ref{figEHFlow} shows the pathway leading to $ \widehat R $ and then finally to the transformed Einstein-Hilbert action, starting from the transformed metric $ \widehat g_{\mu \nu } $ and its inverse $ \widehat g^{\mu \nu } $.  

The transformed metric and its inverse metric  are needed for the calculation of the transformed Christoffel symbol $ \widehat \Gamma ^{\alpha }_{\mu \nu } $.
\begin{align}
	\widehat \Gamma ^\alpha _{\mu \nu }
	&=
	\frac{1}{2}\widehat g^{\alpha \beta }\big(
		\widehat g_{\beta \mu ,\nu }
		+
		\widehat g_{\nu \beta ,\mu }
		-
		\widehat g_{\mu \nu ,\beta }
	\big).	
\end{align}
With the Christoffel symbol in hand, the transformed Ricci tensor $ \widehat R_{\mu \nu } $ can be computed from the contraction of the transformed Riemann curvature tensor $ \widehat R^\alpha {}_{\mu \beta \nu } $, which depends on $ \widehat \Gamma ^{\alpha }_{\mu \nu } $ and its derivatives.
\begin{align}
	\widehat R^\alpha {}_{\mu \beta \nu }
	&=
	-\widehat \Gamma ^\alpha _{\mu \beta ,\nu }
	+
	\widehat \Gamma ^\alpha _{\mu \nu ,\beta }
	-
	\widehat \Gamma ^\rho _{\mu \beta }\widehat \Gamma ^\alpha _{\rho \nu }
	+
	\widehat \Gamma ^\rho _{\mu \nu }\widehat \Gamma ^\alpha _{\rho \beta }.
\end{align}
Once $ \widehat R_{\mu \nu } $ is known, the transformed Ricci scalar $ \widehat R $ can be determined through the relation $ \widehat R = \widehat g^{\mu \nu }\widehat R_{\mu \nu } $.  This then, together with the expression for the transformed metric determinant $ \widehat g $, leads to the transformed Einstein-Hilbert action.

\end{multicols}

\begin{multicols}{2}

\subsection{Metric Inverse and its Determinant}
Consider a square matrix \textbf{M} of dimensions $ n\times n $, where $ n $ is a natural number and two column matrices \textbf{u} and \textbf{v} both with dimensions $ n \times 1 $. The inverse of the sum $ \mathbf M + \mathbf u\mathbf v^T$, where the superscript $ T $ denotes transposition, is given by the \textit{Sherman-Morrison formula}:
\begin{align}
	(\mathbf{M} + \mathbf{u}\mathbf{v}^T)^{-1}
	=
	\mathbf {M}^{-1}
	-
	\frac{\mathbf {M}^{-1}\mathbf {u} \mathbf {v}^T\mathbf {M}^{-1}}{
		1 + \mathbf {v}^T\mathbf {M}^{-1}\mathbf {u}
	}.
\end{align}
For the disformally transformed metric, $ \widehat g_{\mu \nu } = A(\phi )g_{\mu \nu } + B(\phi )\phi _{;\mu }\phi _{;\nu } $, we may take $ \mathbf{M} = \{Ag_{\mu \nu }\},\, \mathbf u = \{B \phi_{;\mu }\},$ and $ \mathbf v = \{\phi _{;\nu}\}$, to calculate for $ \widehat g^{\mu \nu } $. It follows that
\begin{align}
	\label{gInv}
	\widehat g^{\mu \nu }
	&=
	\frac{g^{\mu \nu }}{A}
	-
	\frac{1}{A^2}
	\frac{Bg^{\mu \alpha}\phi _{;\alpha }\phi _{;\beta }g^{\beta \nu }}{
		1 + (B/A)g^{\mu \nu }\phi_{;\mu }\phi _{;\nu }
	},
	\nonumber
	\\[0.5em]
	&\hspace{-2.0em}\boxed{\phantom{a}
		\widehat g^{\mu \nu }
		=
		\frac{1}{A}\left(
			g^{\mu \nu }
			-
			\frac{B}{A - 2BX}\phi ^{;\mu }\phi ^{;\nu }
		\right),
		\phantom{a}
	}		
\end{align}
where $ X \equiv -\frac{1}{2} g^{\mu \nu }\phi _{;\mu }\phi _{;\nu } $ and the raised covariant derivative index means $ \phi ^{;\mu } = g^{\mu \alpha }\phi _{;\alpha }$. Note that for the inverse to exist, we must impose the requirements $ A \ne 0 $ and $ A - 2BX \ne 0 $. Furthermore, to preserve the metric signature, the conditions given by $ A>0 $ and $ A - 2BX > 0 $, must hold \cite{Bettoni:2013diz}.

The determinant of the disformally transformed metric can be determined by using the equation given by 
\begin{align}
	\det(\mathbf M + \mathbf u \mathbf v^T)
	=
	(1 + \mathbf v^T\mathbf M^{-1}\mathbf u)
	\det(\mathbf M),
\end{align}
which is a consequence of \textit{Silvester's determinant theorem}. Applying this relation for $\{\widehat g_{\mu \nu }\} $ we have for $ \det\{\widehat g_{\mu \nu }\} = \widehat g $,
\begin{align}
	\det\{\widehat g_{\mu \nu }\}
	&=
	\left(
		1
		+
		\frac{B}{A}
		\phi _{;\nu }g^{\mu \nu }
		\phi _{;\mu }
	\right)\det\{Ag_{\mu \nu }\},
	\nonumber
	\\[0.5em]
	\widehat g
	&=
	A^4\left(
		1
		-
		2\frac{B}{A}X
	\right)g.
\end{align}
For the transformed integral measure in the Einstein-Hilbert action, we have
\begin{align}
	\boxed{\phantom{a}
		\sqrt{-\widehat g}\,\dd^4 x
		=
		A^\frac{3}{2}(A - 2BX)^\frac{1}{2}\sqrt{-g}\,\dd^4 x
		=
		D_3\sqrt{-g}\,\dd^4 x,
		\phantom{a}
	}
\end{align}
where we have denoted $ D_3 = A^\frac{3}{2}(A - 2BX)^\frac{1}{2} $. The condition stated above, $ A - 2BX > 0$, further ensures that the integral measure remains real.

\bigskip
\subsection{Christoffel Symbol}
The transformed Christoffel symbol, $ \widehat \Gamma ^\alpha _{\mu \nu } $, follows the same formula as that of the original Christoffel symbol, but the inverse metric and metric (derivatives) now ``wear'' a hat.
\begin{align}
	\widehat \Gamma ^\alpha _{\mu \nu }
	&=
	\frac{1}{2}\widehat g^{\alpha \beta }\big(
		\widehat g_{\beta \mu ,\nu }
		+
		\widehat g_{\nu \beta ,\mu }
		-
		\widehat g_{\mu \nu ,\beta }
	\big)
\end{align}
Here, comma denotes partial differentiation; \textit{i.e.}, $ \phi _{,\nu } = \partial \phi /\partial x^\nu  $. Because $ \widehat g^{\alpha \beta } = g^{\mu \nu }/A + \cdots $ and $ \widehat g_{\mu \nu } = Ag_{\mu \nu } + \cdots $ according to (\ref{disftransf}) and (\ref{gInv}), respectively, we can express $ \widehat \Gamma ^\alpha _{\mu \nu } $ in terms of the original Christoffel symbol $ \Gamma ^\alpha _{\mu \nu } $ plus some tensor terms. We have upon substitution from (\ref{disftransf}) and (\ref{gInv}), and expanding the partial derivatives,
\begin{align}
	\widehat \Gamma ^\alpha _{\mu \nu }
	&=
	\frac{1}{2A}\left(
		g^{\alpha \beta }
		-
		\frac{B}{A - 2BX}\phi ^{;\alpha  }\phi ^{;\beta }
	\right)		
	\nonumber
	\\[0.5em]
	&\quad
	\times
	\big(
		A_{;\nu }g_{\beta \mu } 
		+
		Ag_{\beta \mu,\nu }
		+ 
		B_{;\nu }\phi _{;\beta }\phi _{;\mu }
		+
		\underline{B\phi _{;\beta,\nu }\phi _{;\mu }}
		\nonumber
		\\[0.5em]
		&\qquad	
		+\,
		\underline{\underline{B\phi _{;\beta}\phi _{;\mu,\nu }}}
		+
		A_{;\mu }g_{\nu \beta } 
		+
		Ag_{\nu \beta,\mu  }
		+ 
		\underline{B_{;\mu }\phi _{;\nu }\phi _{;\beta }}
		\nonumber
		\\[0.5em]
		&\qquad
		+\, 
		\underline{\underline{B\phi _{;\nu,\mu  }\phi _{;\beta }}}
		+ 
		\underline{B\phi _{;\nu }\phi _{;\beta,\mu  }}				
		-
		A_{;\beta }g_{\mu \nu }
		-
		Ag_{\mu \nu,\beta  }
		\nonumber
		\\[0.5em]
		&\qquad
		-\, 
		\underline{B_{;\beta }\phi _{;\mu  }\phi _{;\nu }}
		- 
		\underline{B\phi _{;\mu,\beta }\phi _{;\nu }}
		- 
		\underline{B\phi _{;\mu  }\phi _{;\nu,\beta }}	
	\big).
\end{align}
Note that $ \phi _{;\mu \nu } = \phi _{;\mu ,\nu } - \Gamma ^\alpha _{\mu \nu }\phi _{;\alpha }$; as such, $ \phi _{;\mu ,\nu } = \phi _{;\nu ,\mu }$. Furthermore, $ B_{;\mu } = B'\phi _{;\mu } $ and $ A_{;\mu } = A'\phi _{;\mu } $, where the prime indicates differentiation with respect to $ \phi  $. It follows that we can cancel all the terms with single underline and combine the pair of double underlined terms above. After expanding, regrouping terms, and identifying the original Christoffel symbol, we find
\begin{align}
	\widehat \Gamma ^\alpha _{\mu \nu }
	&=
	\Gamma^\alpha _{\mu \nu }
	+
	\frac{A'}{2A}\big(
		\phi _{;\nu }\delta ^\alpha _\mu 
		+
		\phi _{;\mu }\delta ^\alpha _\nu 
	\big)
	-
	\frac{A'}{2(A - 2BX)}\phi ^{;\alpha }g_{\mu \nu }
	\nonumber
	\\[0.5em]
	&\quad	
	+\,
	\frac{AB' - 2A'B}{2A(A - 2BX)}
	\phi ^{;\alpha }\phi _{;\mu }\phi _{;\nu }
	\\[0.5em]
	&\quad
	+\,
	\frac{B}{2(A - 2BX)}\phi ^{;\alpha }\big(
		\phi ^{;\beta }{}_{,\nu }g_{\beta \mu }
		+
		\phi ^{;\beta }{}_{,\mu }g_{\nu \beta }
		+
		\phi ^{;\beta }g_{\mu \nu ,\beta }
	\big).
	\nonumber
\end{align}
The Kronecker delta emerged from the metric contraction; e.g., $ g^{\alpha \beta }g_{\beta \mu } = \delta ^{\alpha }_\mu  $. 

Observe that the last term in the last equation above for $ \widehat \Gamma ^\alpha _{\mu \nu } $ involves partial derivative terms. From the equation for covariant derivative, 
we can write $ \phi ^{;\beta }{}_{,\nu } = \phi ^{;\beta }{}_{;\nu } - \Gamma ^\beta _{\nu \rho }\phi ^{;\rho } $. It follows that
\begin{align}
	&\phi ^{;\beta }{}_{,\nu }g_{\beta \mu }
	+
	\phi ^{;\beta }{}_{,\mu }g_{\nu \beta }
	\nonumber
	\\[0.5em]
	&\quad=
	\big(
		\phi ^{;\beta }{}_{;\nu } - \Gamma ^\beta _{\nu \rho }\phi ^{;\rho }
	\big)g_{\beta \mu }
	+
	\big(
		\phi ^{;\beta }{}_{;\mu } - \Gamma ^\beta _{\mu \rho }\phi ^{;\rho }
	\big)g_{\beta \nu },
	\nonumber
	\\[0.35em]
	&\quad=
	2\phi _{;\mu \nu }
	-
	\phi ^{;\rho }\big(
		\Gamma_{\mu \nu \rho }
		+
		\Gamma_{\nu \mu \rho }
	\big),
	\nonumber
	\\[0.35em]
	&\quad=
	2\phi _{;\mu \nu }
	-
	\tfrac{1}{2}\,\phi ^{;\rho }\big(
		g_{\mu \nu ,\rho }
		+
		g_{\rho \mu ,\nu }
		-
		g_{\nu \rho ,\mu }
		\nonumber
		\\[0.5em]
		&\qquad\qquad\quad
		+\,
		g_{\nu \mu ,\rho }
		+
		g_{\rho \nu ,\mu }
		-
		g_{\mu \rho ,\nu }
	\big),
	\nonumber
	\\[0.35em]
	&\quad=
	2\phi _{;\mu \nu }
	-
	\phi ^{;\rho }	g_{\mu \nu ,\rho }.
\end{align}
Substitution in the equation above for $ \widehat \Gamma ^\alpha _{\mu \nu } $ yields our sought-for expression for the transformed Christoffel symbol.
\begin{equation}
	\label{disforChris}
	\boxed{\phantom{a}
		\begin{aligned}
			\widehat \Gamma ^\alpha _{\mu \nu }
			&=
			\Gamma^\alpha _{\mu \nu }
			+
			\frac{A'}{2A}\big(
				\phi _{;\nu }\delta ^\alpha _\mu 
				+
				\phi _{;\mu }\delta ^\alpha _\nu 
			\big)
			\nonumber
			\\[0.5em]
			&\qquad
			-\,
			\frac{A'}{2(A - 2BX)}\phi ^{;\alpha }g_{\mu \nu }
			\\[0.5em]
			&\qquad
			+\,
			\frac{AB' - 2A'B}{2A(A - 2BX)}
			\phi ^{;\alpha }\phi _{;\mu }\phi _{;\nu }
			\nonumber
			\\[0.5em]
			&\qquad
			+\,
			\frac{B}{(A - 2BX)}\phi ^{;\alpha }\phi _{;\mu\nu  }.
		\end{aligned}
		\phantom{a}
	}
\end{equation}
We remark that the difference $ \widehat \Gamma ^\alpha _{\mu \nu } - \Gamma ^\alpha _{\mu \nu }$ is a tensor although $ \widehat \Gamma ^\alpha _{\mu \nu } $ and $ \Gamma ^\alpha _{\mu \nu } $ are both non-tensorial in nature. Moreover, $ \widehat \Gamma ^\alpha _{\mu \nu } $ retains the symmetry with respect to the lower indices $ (\mu ,\nu ) $ as in $ \Gamma ^\alpha _{\mu \nu } $.

\bigskip
\subsection{Ricci Tensor and Ricci Scalar}
Given the transformed Riemann curvature tensor $ \widehat R^\alpha {}_{\mu \beta \nu } $, the transformed Ricci tensor is simply the contraction of the indices $ (\alpha ,\beta)  $; that is, $ \widehat R_{\mu \nu } = \widehat R^\alpha {}_{\mu \alpha \nu } $. On the other hand, 
\begin{align}
	\widehat R^\alpha {}_{\mu \beta \nu }
	&=
	-\widehat \Gamma ^\alpha _{\mu \beta ,\nu }
	+
	\widehat \Gamma ^\alpha _{\mu \nu ,\beta }
	-
	\widehat \Gamma ^\rho _{\mu \beta }\widehat \Gamma ^\alpha _{\rho \nu }
	+
	\widehat \Gamma ^\rho _{\mu \nu }\widehat \Gamma ^\alpha _{\rho \beta }.
\end{align}
If we let $ \widehat \Gamma ^\alpha _{\mu \nu } = \Gamma ^\alpha _{\mu \nu } + C ^\alpha _{\mu \nu } $ in (\ref{disforChris}) where 
\begin{align}
	\label{CAlpMuNu}
	C ^\alpha _{\mu \nu }
	&=
	\frac{A'}{2A}\big(
		\phi _{;\nu }\delta ^\alpha _\mu 
		+
		\phi _{;\mu }\delta ^\alpha _\nu 
	\big)
	-
	\frac{A'}{2(A - 2BX)}\phi ^{;\alpha }g_{\mu \nu }
	\nonumber
	\\[0.35em]
	&\qquad
	+\,
	\frac{AB' - 2A'B}{2A(A - 2BX)}
	\phi ^{;\alpha }\phi _{;\mu }\phi _{;\nu }
	\nonumber
	\\[0.5em]
	&\qquad
	+\,
	\frac{B}{(A - 2BX)}\phi ^{;\alpha }\phi _{;\mu\nu  },
\end{align}
then
\begin{align}
	\widehat R^\alpha {}_{\mu \beta \nu }
	&=
	R^\alpha {}_{\mu \beta \nu }
	-
	{C^\alpha _{\mu \beta ,\nu }}
	+
	{C^\alpha _{\mu \nu ,\beta }}
	-
	{\Gamma ^\alpha _{\rho \nu }C^\rho _{\mu \beta }}
	-
	{\Gamma ^\rho _{\mu \beta }C^\alpha _{\rho \nu }}
	\nonumber
	\\[0.5em]
	&\quad
	-\,
	C^\rho _{\mu \beta }C^\alpha _{\rho \nu }
	+
	{\Gamma ^\alpha _{\rho \beta }C^\rho _{\mu \nu }}
	+
	{\Gamma ^\rho _{\mu \nu }C^\alpha _{\rho \beta }}
	+
	C^\rho _{\mu \nu }C^\alpha _{\rho \beta }.
\end{align}

The right hand side involves partial derivative terms. The tensorial nature of $ \widehat R^\alpha {}_{\mu \beta \nu } $ can be made manifest by noting that 
\begin{align}
	C^\alpha _{\mu \beta ;\nu }
	&=
	C^\alpha _{\mu \beta ,\nu }
	+
	\Gamma ^\alpha _{\nu \rho }C^\rho _{\mu \beta }
	-
	\Gamma ^\rho _{\mu \nu }C^\alpha _{\rho \beta }
	-
	\Gamma ^\rho _{\beta \nu }C^\alpha _{\mu \rho }
	\nonumber
	\\[0.5em]
	C^\alpha _{\mu\nu;\beta}
	&=
	C^\alpha _{\mu\nu,\beta }
	+
	\Gamma ^\alpha _{\beta \rho }C^\rho _{\mu \nu }
	-
	\Gamma ^\rho _{\mu\beta}C^\alpha _{\rho \nu }
	-
	\Gamma ^\rho _{\nu \beta }C^\alpha _{\mu \rho }.
\end{align}
Upon substitution of this pair of expressions in the equation above for $ \widehat R^\alpha {}_{\mu \beta \nu }  $ and contracting the indices $ (\alpha ,\beta ) $ we find
\begin{align}
	\label{hatRmunu}
	\widehat R_{\mu \nu }
	&=
	\widehat R^\alpha {}_{\mu \alpha \nu },
	\\[0.5em]
	&=
	R_{\mu \nu }
	-
	C^\alpha _{\mu \alpha ;\nu }
	+
	C^\alpha _{\mu \nu ;\alpha }
	-
	C^\rho _{\mu \alpha }C^\alpha _{\rho \nu }
	+
	C^\rho _{\mu \nu }C^\alpha _{\rho \alpha },
	\nonumber
\end{align}
which is now clearly a tensorial equation. We see that $ \widehat R_{\mu \nu } $ retains the symmetry with respect to its pair of indices as in $ R_{\mu \nu } $. The terms following $ R_{\mu \nu } $ are due to the disformal transformation. They are functionals of $ X  $ and $ \phi  $ with the later being encoded in the conformal factor $ A(\phi ) $ and disformal factor $ B(\phi ) $.

The calculation of the terms involving $ C^\alpha _{\mu \nu } $ in the equation above for $ \widehat R_{\mu \nu } $ takes some amount of space, but for the most part, straightforward. For the sum of the last two terms, namely, $-C^\rho _{\mu \alpha }C^\alpha _{\rho \nu } +	C^\rho _{\mu \nu }C^\alpha _{\rho \alpha } $, we have upon using (\ref{CAlpMuNu}) for $ C^\alpha _{\mu \nu } $,
\begin{align}
	\label{secCC}
	&-C^\rho _{\mu \alpha }C^\alpha _{\rho \nu } + C^\rho _{\mu \nu }C^\alpha _{\rho \alpha}
	\nonumber
	\\[0.5em]
	&\qquad
	=
	\frac{A'g_{\mu \nu }}{2A(A - 2BX)^2}
	\big[
		A(
			BX_{;\alpha }\phi^{;\alpha }
			+ 
			2A'X
		)
		\nonumber
		\\[0.5em]
		&\qquad\qquad\qquad
		-\,
		2X^2(
			A'B + AB'
		)
	\big]
	\nonumber
	\\[0.5em]
	&\qquad\qquad
	-\,
	\frac{\phi _{;\mu }\phi _{;\nu }}{2A^2(A - 2BX)^2}\big[
		6A'BX^2(A'B - AB')
		\nonumber
		\\[0.5em]
		&\qquad\qquad\qquad
		-\,
		2AA'X(
			A'B - 2B'A
		)
		\\[0.5em]
		&\qquad\qquad\qquad
		-\,
		ABX_{;\alpha }\phi ^{;\alpha }(
			2A'B
			-
			AB'
		)
		-
		A^2A'^2	
	\big]
	\nonumber
	\\[0.5em]
	&\qquad\qquad
	-\,
	\frac{B(\phi _{;\mu }X_{;\nu } + X_{;\mu }\phi _{;\nu })}{2A(A - 2BX)^2}\big[
		2X(AB' - 2A'B) + A'A
	\big]
	\nonumber
	\\[0.5em]
	&\qquad\qquad
	+
	\frac{B\phi _{;\mu \nu }}{A(A - 2BX)^2}\big[
		2X^2(A'B + AB')
		\nonumber
		\\[0.5em]
		&\qquad\qquad\qquad
		-\,
		A(2A'X + X_{;\alpha }\phi^{;\alpha }B)
	\big]	
	-
	\frac{B^2X_{;\mu }X_{;\nu }}{(A - 2BX)^2}
	\nonumber
\end{align}
We observe that all the terms on the right hand side are symmetric with respect to the indices $ (\mu ,\nu ) $, consistent with the symmetry of $ \widehat R_{\mu \nu } $.

The first two terms following $ R_{\mu \nu } $ in (\ref{hatRmunu}), namely, $ C^\alpha _{\mu \alpha ;\nu } $ and $ C^\alpha _{\mu \nu ;\alpha } $, require some attention because of the existence of third order derivative terms. We have
\begin{align}
	\label{firstCC}
	C ^\alpha _{\mu \alpha;\nu }
	&=
	\frac{\phi _{;\mu }\phi _{;\nu }}{A^2(A - 2BX)^2}\big(
		2A^2BB'' X^2
		-
		2A^2{B' }^2X^2
		\nonumber
		\\[0.5em]
		&\qquad
		+\,
		6AA'' B^2X^2
		-
		6A'^2B^2X^2
		-
		A^{3}B'' X
		\nonumber
		\\[0.5em]
		&\qquad
		+\,
		2A^2A' B' X
		-
		7A^2A'' BX
		+
		6AA'^2BX
		\nonumber
		\\[0.5em]
		&\qquad
		+\,
		2A^3A'' 
		-
		2A^2A'^2
	\big)
	\nonumber
	\\[0.5em]
	&\quad
	-\,
	\frac{X_{;\mu }\phi _{;\nu } + \phi _{;\mu }X_{;\nu }}{(A - 2BX)^2}\big(
		AB' - A'B
	\big)
	\nonumber
	\\[0.5em]
	&\quad
	-\,
	\frac{2B^2}{(A - 2BX)^2}X_{;\mu }X_{;\nu }	
	\nonumber
	\\[0.5em]
	&\quad
	+\,
	\frac{2A'A - (AB' + 3A'B)X}{A(A - 2BX)}\phi _{;\mu\nu}
	\nonumber
	\\[0.5em]
	&\quad
	-\,
	\boxed{\frac{B}{A - 2BX}X_{;\mu\nu }},
	\nonumber
	\\[0.5em]
	C ^\alpha _{\mu \nu;\alpha }
	&=
	\frac{2A'B - AB'}{2A(A - 2BX)} (
		\phi _{;\mu }X_{;\nu }
		+
		\phi _{;\nu  }X_{;\mu  }
	)
	\nonumber
	\\[0.5em]
	&\quad
	-\,
	\frac{2A'B - AB'}{2A(A - 2BX)} 
	(\square \phi)\phi _{;\mu }\phi _{;\nu }
	\nonumber
	\\[0.5em]
	&\quad
	-\,
	\frac{B(2A'B - AB')}{A(A - 2BX)^2}
	(X^{;\alpha }\phi _{;\alpha })	
	\phi _{;\mu }\phi _{;\nu }
	\nonumber
	\\[0.5em]
	&\quad
	+\,
	\frac{\phi _{;\mu }\phi _{;\nu }}{(A - 2BX)^2} \big[
		2X^2(BB'' - B'^2)
		\nonumber
		\\[0.5em]
		&\qquad
		-\,
		X(AB'' - 3A'B' + 2A''B)
		+
		AA'' - A'^2		
	\big]
	\nonumber
	\\[0.5em]
	&\quad
	+\,
	\boxed{
	\frac{B\phi ^{;\alpha }\phi _{;\mu \nu \alpha }}{A - 2BX}
	}
	-
	\frac{A'(\square \phi )g_{\mu \nu } }{2(A - 2BX)}
	\nonumber
	\\[0.5em]
	&\quad
	-\,
	\frac{A'B(X^{;\alpha }\phi _{;\alpha })g_{\mu \nu }}{(A - 2BX)^2}
	\nonumber
	\\[0.5em]
	&\quad
	+\,
	\frac{Xg_{\mu \nu }}{(A - 2BX)^2}\big[
		2X(A'B' - A''B)
		\nonumber
		\\[0.5em]
		&\qquad
		+\,
		AA'' - A'^2
	\big]
	+
	\frac{B(\square \phi )\phi _{;\mu \nu }}{A - 2BX}
	\nonumber
	\\[0.5em]
	&\quad
	+\,
	\frac{2B^2(X^{;\alpha }\phi _{;\alpha })\phi _{;\mu \nu }}{
		(A - 2BX)^2
	}
	+
	\frac{\phi _{;\mu \nu }}{A(A - 2BX)^2}\big[
		4A'B^2X^2
		\nonumber
		\\[0.5em]
		&\qquad
		-\,
		2AX(AB' + A'B)
		+
		A^2A'
	\big].
\end{align}
Because $ X \equiv -\frac{1}{2} g^{\mu \nu }\phi _{;\mu }\phi _{;\nu } $, the derivative term $ X_{;\mu \nu } $ involves third-order derivative of $ \phi $. However, the combination, $ -C^\alpha _{\mu \alpha ;\nu } + C^\alpha _{\mu \nu ;\alpha } $, eliminates these third-order derivative terms through the definition of the Riemann curvature tensor; in particular, given a tensor $ [\nabla _\mu ,\nabla _\nu ]V^\alpha = R^\alpha {}_{\beta \mu \nu }V^\beta $. We have
\begin{align}
	\label{XmnPmna}
	X_{;\mu \nu }
	+
	\phi ^{;\alpha }\phi _{;\mu \nu \alpha }
	&=
	\phi ^{;\alpha }\big(
		\phi _{;\mu \nu \alpha }
		-
		\phi _{;\mu\alpha\nu }
	\big)
	-
	\phi _{;\alpha \mu }\phi ^{;\alpha }{}_{;\nu },
	\nonumber
	\\[0.5em]
	X_{;\mu \nu }
	+
	\phi ^{;\alpha }\phi _{;\mu \nu \alpha }
	&=
	-\phi ^{;\alpha }\phi ^{;\beta }
	R_{\mu \alpha \nu \beta }
	-
	\phi _{;\alpha \mu }\phi ^{;\alpha }{}_{;\nu },
\end{align}
which no longer explicitly involves third-order derivative of $ \phi  $.

Having settled the third order derivative terms, we perform substitution from (\ref{firstCC}) and (\ref{secCC}) in the equation for the hatted Ricci tensor given by (\ref{hatRmunu}). We find
\begin{align}
	\widehat R_{\mu \nu }
	&=
	R_{\mu \nu }
	-
	D_1B \phi ^{;\alpha }\phi ^{;\beta  }R_{\mu \alpha \nu \beta }
	+
	D_1^2D_2X\big[
		(AA'B'
		\nonumber
		\\[0.5em]
		&\qquad
		-\, 
		2AA''B - A'^2B)X
		+
		A^2A''
	\big] g_{\mu \nu }	
	\nonumber
	\\[0.5em]
	&\quad 
	-\,
	\tfrac{1}{2}D_1A'(\square \phi) g_{\mu \nu }
	-
	\tfrac{1}{2}D_1^2 A'B \,\phi ^{;\alpha }X_{;\alpha } g_{\mu \nu }
	\nonumber
	\\[0.5em]
	&\quad
	+
	\tfrac{1}{2}D_1^2D_2^2\phi _{;\mu }\phi _{;\nu } \big[
		6BX^2(AA'B' - 2AA''B + A'^2B)
		\nonumber
		\\[0.5em]
		&\qquad
		-\,
		2AX(AA'B'		
		-
		5AA''B + 5A'^2B)
		\nonumber
		\\[0.5em]
		&\qquad 
		-\,
		A^2(2AA'' - 3A'^2)
	\big]
	-
	D_1B\phi ^{;\alpha }{}_\mu  \phi_{;\alpha \nu }
	\nonumber
	\\[0.5em]
	&\quad
	+\,
	\tfrac{1}{2}D_1D_2(AB' - 2A'B)
	(\square \phi)\phi _{;\mu }\phi _{;\nu }
	\nonumber
	\\[0.5em]
	&\quad
	+\,
	\tfrac{1}{2}D_1^2D_2B(AB' - 2A'B)
	\phi^{;\alpha }X_{;\alpha }\phi _{;\mu }\phi _{;\nu }
	\nonumber
	\\[0.5em]
	&\quad 
	-\,
	D_1^2\big[(AB' - 3A'B)X + AA'\big]\phi _{;\mu \nu } 
	\nonumber
	\\[0.5em]
	&\quad 
	+\,
	D_1B(\square \phi )\phi _{;\mu \nu }	
	+
	D_1^2B^2\,\phi ^{;\alpha }X_{;\alpha }\phi _{;\mu \nu }
	\\[0.5em]
	&\quad
	+\,
	D_1^2(AB' - A'B)\phi _{(;\mu }X_{;\nu )}
	+
	D_1^2B^2X_{;\mu }X_{;\nu },
	\nonumber
\end{align}
where
\begin{align}
	D_1 &\equiv \frac{1}{A - 2BX},
	\quad 
	D_2 \equiv \frac{1}{A},
	\nonumber
	\\[0.5em]
	X_{(;\mu }\phi _{;\nu )}
	&=
	\frac{1}{2}(X_{;\mu }\phi _{;\nu } + X_{;\nu }\phi _{;\mu }) 
\end{align}

To get the Ricci scalar, all that is left now is to contract $ \widehat g^{\mu \nu } $ and $ \widehat R_{\mu \nu } $. We have after some re-arrangement, grouping, contraction, and using the definition $ X \equiv -\frac{1}{2} g^{\mu \nu }\phi _{;\mu }\phi _{;\nu } $, 
\begin{equation}
	\boxed{
	\phantom{a}
	\begin{aligned}
	\widehat R
	&=
	D_2R
	-
	2D_1D_2B\,\phi ^{;\mu }\phi ^{;\nu }R_{\mu \nu }
	\\[0.5em]
	&\quad 
	-\,
	D_1^2D_2\big[
		2X(AB' - 4A'B) + 3AA'
	\big]\square \phi  	
	\\[0.5em]
	&\quad 
	+\,
	D_1D_2B\big[
		(\square \phi )^2 - \phi ^{;\mu \nu }\phi _{;\mu \nu }
	\big]
	\nonumber
	\\[0.5em]
	&\quad
	+\,
	2D_1^2D_2B^2\,\phi ^{;\alpha }X_{;\alpha }\,
	\square \phi
	\nonumber
	\\[0.5em]
	&\quad
	+\,
	D_1^2D_2(AB' - 4A'B)\phi ^{;\alpha }X_{;\alpha } 
	\\[0.5em]
	&\quad
	+\,
	2D_1^2D_2B^2\,X^{;\alpha }X_{;\alpha },
	\nonumber
	\\[0.5em]
	&\quad
	+\,
	3D_1^2D_2X\big[
		2X(A'B' - 2A''B) + 2AA'' - A'^2	
	\big].
	\end{aligned}
	\phantom{a}	
	}
\end{equation}
The transformed Ricci scalar is a sum of the original Ricci scalar coupled with the inverse of the conformal factor, $ D_2R $, and terms attributable to the disformal transformation. These additional terms involve $ A $ and $ B $, and fundamentally depends on $ (\phi,X) $ and their derivatives. When $ B $ is set to vanish in the disformal transformation, the equation above reduces to 
\begin{align}
	\label{confricsca}
	\widehat R
	&=
	\frac{R}{A}
	+
	\frac{3X}{A^3}\big(
		2AA'' - A'^2
	\big)
	-
	\frac{3A'}{A^2}\square \phi,
\end{align}
which is the familiar conformal transformation of the Ricci scalar \cite{Wald:1984rg}.

\bigskip
\section{Connection with the Horndeski Action and Stability}
\label{secConnHornd}
The conformal transformation of the Einstein-Hilbert action leads to an action with corresponding equations of motion for $ \phi $ that involve up to second order derivatives only\footnote{The equation of motion for the metric is the well-known Einstein field equations which involve second derivatives of $ g_{\mu \nu } $}. This implies stability. Classically, the Ostrogradsky theorem \cite{Woodard:2015zca} tells us that if the equation of motion is up to second order only, the energy (corresponding to the Hamiltonian) is bounded from below. From the perspective of quantum field theory, it means the absence of (unphysical) ghost fields. Higher order equations motion (greater than two) requires for the action, constraints which are usually hard to find, to remove physically undesirable degrees of freedom. \cite{Faddeev:2010zz,Gleyzes:2014qga,Langlois:2015cwa,Achour:2016rkg}. 

The equations for $ \widehat R $ and $ \sqrt{-\widehat g}\,\dd^4 x  $ in the preceding section, lead directly to the disformally transformed Einstein-Hilbert action given by 
\begin{align}
	\label{transfEHAct}
	\widehat S
	&=
	\int \sqrt{-\widehat g}\,\dd^4x\,\widehat R,
	\nonumber
	\\[0.5em]
	\widehat S
	&=
	\int D_3\sqrt{-g}\,\dd^4 x\, \Big\{D_2R
	-
	2D_1D_2B\,\phi ^{;\mu }\phi ^{;\nu }R_{\mu \nu }
	\nonumber
	\\[0.5em]
	&\qquad
	-\,
	D_1^2D_2\big[
		2X(AB' - 4A'B) + 3AA'
	\big]\square \phi  
	\nonumber
	\\[0.5em]
	&\qquad
	+\,
	D_1D_2B\big[
		(\square \phi )^2 - \phi ^{;\mu \nu }\phi _{;\mu \nu }
	\big]
	\nonumber
	\\[0.5em]
	&\qquad
	+\,
	2D_1^2D_2B^2\,\phi ^{;\alpha }X_{;\alpha }\,
	\square \phi 
	\nonumber
	\\[0.35em]
	&\qquad
	+\,
	D_1^2D_2(AB' - 4A'B)\phi ^{;\alpha }X_{;\alpha }
	\nonumber
	\\[0.5em]
	&\qquad
	+\,
	2D_1^2D_2B^2\,X^{;\alpha }X_{;\alpha }
	\\[0.2em]
	&\qquad
	+\,
	3D_1^2D_2X\big[
		2X(A'B' - 2A''B) + 2AA'' - A'^2	
	\big]
	\Big\}.
	\nonumber
\end{align}
Contained in this transformed action are terms that could lead to equations of motion for $ \phi $ that are higher than second order in derivatives. To be clear, all derivative terms above involving $ \phi  $ are at most second order. But as the Euler-Lagrange equation involves first and second order covariant differentiations, question may arise about the stability of the transformed action above. Similar to that for the conformally transformed Einstein-Hilbert action, this can be addressed for the field $ \phi $ by solving the equations of motion to explicitly see if they are at most second order in nature. Alternatively, we may attempt to ``match'' $ \widehat S $ to the Horndeski action \cite{Horndeski:1974wa}; this is in fact, what we do below. Knowing that this action is the most general scalar tensor action involving a single scalar field that yields at most second order equations of motion, if we can show that the transformed action above is a special case of the Horndeski action, then that effectively solves the (possible) stability issue.

One thing to note at this point is that the Horndeski action is composed of four sub-Lagrangians and the entire action is form-invariant \cite{Bettoni:2013diz} under the disformal transformation given by (\ref{disftransf}). In particular, the Horndeski action $ S_H $ is given by 
\begin{align}
	\label{hornAct}
	S_H
	&=
	\int \dd^4 x\,\sqrt{-g} \,\big(
		\mathcal L_2 + \mathcal L_3
		+
		\mathcal L_4 + \mathcal L_5
	\big),
\end{align}
where
\begin{align}
	\label{hornActL}
	\mathcal L_2
	&=
	P,
	\qquad
	\mathcal L_3
	=
	-G_3\,\square\phi,
	\nonumber
	\\[0.5em]
	\mathcal L_4
	&=
	G_4R
	+ 
	G_{4,X}[(\square \phi )^2 - \phi _{;\mu \nu }\phi ^{;\mu \nu }],
	\nonumber
	\\[0.5em]
	\mathcal L_5
	&=
	G_5G_{\mu \nu }\phi ^{;\mu \nu }
	-
	\frac{1}{3!}G_{5,X}[
		(\square\phi )^3
		-
		3(\square\phi )\phi _{;\mu \nu }\phi ^{;\mu \nu }
		\nonumber
		\\[0.5em]
		&\qquad	\qquad\qquad\qquad\qquad
		+\,
		2\phi _{;\mu \alpha }\phi ^{;\alpha \nu }\phi ^{;\mu }{}_{\nu }
	],
\end{align}
with $ P,\, G_3,\, G_4, $ and $ G_5 $ being functionals of $ (X, \phi ) $,\, $ \square \phi = g^{\mu \nu }\phi _{;\mu \nu } $, and $ G_{\mu \nu } $ is the Einstein tensor. The disformal transformation of each term in the equation for $ S_H $ may generate third order derivative terms. Nevertheless, in combination, third order terms are cancelled or absorbed through the Riemann curvature tensor, and the transformed action (by regrouping the transformed terms,) retains the same form as the original Horndeski action.

For the case at hand, we have only \textit{one} term to consider, namely, the Einstein-Hilbert Lagrangian. It is a special case of $ \mathcal L_4 $ where $ G_4 = 1 $. With no ``neighbouring'' term in sight to possibly help cancel some problematic terms in the transformation as in the case of $ \mathcal L_5 $\footnote{The pair of terms in $ \mathcal L_5 $ work in tandem to cancel problematic terms in $ \widehat {\mathcal L}_5 $; the same goes for the pair of terms in $ \widehat {\mathcal L}_4 $, for instance, given a general form of $ G_4 = G_4(\phi ,X) $.}, it is interesting to know if the Einstein-Hilbert action has the ``special'' feature of yielding a stable transformed action under the disformal transformation. We wish to demonstrate that this is indeed, the case. As we have  mentioned above, the calculation shall be by way of ``matching'' it to the Horndeski action as the latter's special case. 

Looking at the transformed Einstein-Hilbert action given by (\ref{transfEHAct}), we see four terms that do not belong to any of the sub-Lagrangians of the Horndeski action in their current form. These terms are 
\begin{align}
	\label{outcsterms}
	&
	-2D_1D_2B\,\phi ^{;\mu }\phi ^{;\nu }R_{\mu \nu }
	+
	2D_1^2D_2B^2\,\phi ^{;\alpha }X_{;\alpha }\square \phi 
	\\[0.35em]
	&\quad 	
	+\,
	2D_1^2D_2B^2\,X^{;\alpha }X_{;\alpha }
	+
	D_1^2D_2(AB' - 4A'B)\phi ^{;\alpha }X_{;\alpha } 
	\nonumber
\end{align}
These terms are multiplied by $ D_3\sqrt{-g}  $ in the transformed action.

For the first term above, we express the Ricci tensor in terms of derivatives of $ \phi  $ and perform integration by parts. Denoting $ D_4 = D_1D_2D_3B $ we have
\begin{align}
	\label{fmufnuRexpd}
	&\int \sqrt{-g} \,\dd^4 x\,\big(
		-2D_4\,\phi ^{;\mu }\phi ^{;\nu }R_{\mu \nu }
	\big)
	\nonumber
	\\[0.5em]
	&=
	\int \sqrt{-g} \,\dd^4 x\,\big(
		-2D_4\,\phi ^{;\nu }\phi ^{;\mu }{}_{\nu \mu }
		+
		2D_4\,\phi ^{;\nu }\phi ^{;\mu }{}_{\mu \nu }
	\big),
	\nonumber
	\\[0.5em]
	&=
	\int \sqrt{-g} \,\dd^4 x\,\big\{
		-2D_{4,\phi }\,\phi ^{;\alpha  }X_{;\alpha }
		-
		2D_{4,X}\,X^{;\alpha  }X_{;\alpha }
		\nonumber
		\\[0.5em]
		&\qquad
		-\,
		2D_4\,[
			(\square \phi )^2
			-
			\phi ^{;\mu \nu }\phi _{;\mu \nu }
		]
		\nonumber
		\\[0.35em]
		&\qquad
		+\,
		4D_{4,\phi }X\square \phi 
		-
		2D_{4,X}\phi ^{;\alpha }X_{;\alpha }\square \phi 
	\big\}	
\end{align}
The second and last terms inside the pair of curly braces above cancel the second and third terms in (\ref{outcsterms}) (multiplied by $D_3$).

The only unaccounted terms are those involving $ \phi ^{;\alpha }X_{;\alpha } $. 
Combining all these terms in the transformed action, we have
\begin{align}
	&-\int \sqrt{-g}\,\dd^4x\,
	D_1^2D_2^2D_3
	\nonumber
	\\[0.5em]
	&\qquad
	\times\,
	\big[
		(A - 2BX)(AB)' + 3AA'B
	\big] \phi ^{;\alpha }X_{;\alpha }.
\end{align}
Let us denote 
\begin{align}
	F
	=
	\int \dd X\, D_1^2D_2^2D_3\big[
		(A - 2BX)(AB)' + 3AA'B
	\big].
\end{align}
Because $ F = F(\phi ,X) $ then $ F_{,X}X_{;\alpha } = F_{;\alpha } - F_{,\phi }\phi _{;\alpha }$. The last integral above involving $\phi ^{;\alpha }X_{;\alpha }$ can then be rewritten as
\begin{align}
	&-\int \sqrt{-g}\,\dd^4x\,
	\big[
		F_{;\alpha }\phi ^{;\alpha } 
		- 
		F_{,\phi }\,\phi ^{;\alpha }\phi _{;\alpha }
	\big]
	\nonumber
	\\[0.5em]
	&=
	\int \sqrt{-g}\,\dd^4x\,\big[
		F(\square\phi )
		-
		2XF_{,\phi }
	\big].
\end{align}

Using this equation and the last of (\ref{fmufnuRexpd}) in the equation for $ \widehat S $ given by the second of (\ref{transfEHAct}), we can rewrite the transformed Einstein-Hilbert action as
\begin{equation*}
	\boxed{
	\begin{aligned}
	\widehat S
	&=
	\int \sqrt{-g} \,\dd^4x\,\Big\{
		D_2D_3 R
		\nonumber
		\\[0.35em]
		&\qquad
		-\,
		D_1D_2D_3B\big[
			(\square\phi )^2 - \phi ^{;\mu \nu }\phi _{;\mu \nu }
		\big]		
		\\[0.35em]
		&\qquad
		+\,
		D_1D_2^2D_3\big[
			2X(AB)' - 3AA'
		\big]\square \phi 
		+
		F\square \phi		
		\\[0.35em]
		&\qquad
		+\,
		3D_1^2D_2D_3X\big[
			2X(A'B' - 2A''B)
			\nonumber
			\\[0.35em]
			&\qquad\qquad
			+\, 
			2AA'' - A'^2
		\big]
		-
		2XF_{,\phi }
	\Big\}.
	\end{aligned}
	}
\end{equation*}
Comparing it with the Horndeski action given by (\ref{hornAct}) and (\ref{hornActL}), we identify the terms in the Lagrangian in $ \widehat S $ above as
\begin{align}
	\mathcal L_2 
	&\subset 
	3D_1^2D_2D_3X\big[
		2X(A'B' - 2A''B) + 2AA'' - A'^2
	\big]
	\nonumber
	\\[0.5em]
	&\qquad
	-\,
	2XF_{,\phi }
	\nonumber
	\\[0.35em]
	\mathcal L_3
	&\subset
	D_1D_2^2\big[
		2X(AB)' - 3AA'
	\big]\square \phi 
	+
	F\square \phi,
	\nonumber
	\\[0.35em]
	\mathcal L_4
	&\subset
	D_2D_3 R
	-
	D_1D_2D_3B\big[
		(\square\phi )^2 - \phi ^{;\mu \nu }\phi _{;\mu \nu }
	\big].	
\end{align}
Note that taking $  D_2D_3 $ as a special case of $G_4$ we find $ (D_2D_3)_{,X} = -D_1D_2D_3B $, justifying the identification for $ \mathcal L_4 $ given by the last line above. Hence, with the transformed Einstein-Hilbert action $ \widehat S $ being a special case of the Horndeski action, all the equations of motion for $ \phi $ are second order in derivative terms; $ \widehat S $ describes a stable scalar tensor theory.

It is worth contemplating whether we could end up with the same conclusion had we started with $ \int \sqrt{-g}\,\dd^4x\, G_4(\phi ,X)R $ instead of simply $ \int \sqrt{-g}\,\dd^4x\, R $---the Einstein-Hilbert action. The former describes a field $ \phi $ nonminimally coupled with gravity. The corresponding Lagrangian is also a special case of $ \mathcal L_4 $ as that of the Lagrangian in the Einstein-Hilbert action. As it turns out, the disformal transformation of $ \int \sqrt{-g}\,\dd^4x\, G_4(\phi ,X)R $ leads to terms involving $ G_{4,X} $ in the process of simplifying the term involving $ \phi ^{;\mu }\phi ^{;\nu }R_{\mu \nu } $, similar to what we have done for the transformed Einstein-Hilbert action above. In particular, we have the extra terms given by 
\begin{align}
	&-2D_1D_2D_3\,G_{4,X}\,B\,X_{;\alpha }X^{;\alpha }
	\nonumber
	\\[0.5em]
	&\quad
	-\,
	2D_1D_2D_3\,G_{4,X}B\,\phi ^{;\alpha }X_{;\alpha }\,\square\phi.
\end{align}
Given that $ G_4 $ is a general functional of $ (\phi ,X) $, these cannot be canceled by considering terms in the transformed $ \int \sqrt{-g}\,\dd^4x\, G_4(\phi ,X)R $ alone; a pair of ``counter terms'' is necessary. Such pair of ``counter terms'' emerges from the transformation of $ \int \sqrt{-g}\,\dd^4x\,G_{4,X}[(\square \phi )^2 - \phi _{;\mu \nu }\phi ^{;\mu \nu }] $. It is thus, fitting\footnote{Needless to say, if $ G_4 = G_4(\phi ) $, then similar to that for the Einstein-Hilbert action, the transformed action describes a stable scalar-tensor theory.} that the fourth sub-Lagrangian in the Horndeski action include $ G_{4,X}[(\square \phi )^2 - \phi _{;\mu \nu }\phi ^{;\mu \nu }] $. 

\section{Concluding Remarks}
\label{seConclude}
We live in a complicated universe and our ever-evolving theories describing it may only asymptotically approach its ``true'' nature. Einstein made a significant leap for humankind when he discovered the general theory of relativity that we eventually used (and is still being used) in describing the dynamics of the Universe. Although we made great strides in understanding it through relativity, there remains big puzzles such as the problem of dark energy and dark matter, that we need to solve. Disformal transformation plays a significant role in scalar tensor theories attempting to solve these problems. These theories involve actions that are generalizations of the Einstein-Hilbert action. Similar to that of conformal transformation, disformal transformation functions to relate one theory to another and serves as a symmetry transformation for physical observables such as the power spectrum for the primordial cosmological perturbations.

While the topic of conformal transformation is (almost) a standard part of mainstream graduate texts on gravity and modern cosmology, its generalization to disformal transformation is yet to penetrate these texts. In this paper, we present a pedagogically oriented derivation of the disformal transformation of the Einstein-Hilbert action. The idea is that while this transformation is more complex in nature, by following the same logic as that for conformal transformation, the transformation of the Einstein-Hilbert action can be calculated, albeit with some problematic terms. These problematic terms, with the trick involving the use of the definition of the Riemann curvature tensor and simple integration by parts, can be dealt with properly leading to an apparently stable action.

We have not directly related disformal transformation and scalar tensor theories with the problem of dark energy and dark matter. But for this, research papers abound. And due to its overly complex nature, the approach of wordy but smooth explanations definitely deserves a book of its own. Here, owing the limit in focus and limit in space, we would like to content ourselves with the idea of extending calculations involving conformal transformation commonly found in mainstream graduate texts to the disformal transformation of the Einstein-Hilbert action. We hope that graduate students aiming to venture into modern cosmology may find this paper illuminating.

\section*{Acknowledgement}
\label{acknow}
This work is funded by the University of the Philippines System Enhanced Creative Work and Research Grant (ECWRG 2018-02-11). The author would like to express his sincere gratitude to Prof. Takahiro Kubota of Osaka University (Japan) for fruitful discussions on disformal transformation in the context of modern cosmology. Our gratitude also goes to Joshwa Ordo\~nez of the University of the Philippines for some corrections pointed out in the equations.

\end{multicols}
\end{document}